\documentclass[aps,prc,twocolumn,showpacs,superscriptaddress]{revtex4-1}
\usepackage{graphics}
\usepackage{epsfig}
\usepackage{color}
\usepackage[colorlinks,linkcolor=blue,citecolor=blue]{hyperref}
\usepackage[cp1251]{inputenc}
\usepackage{amssymb,latexsym,amsmath,bm}

\newcommand{\beq}{\begin{equation}}
\newcommand{\eeq}{\end{equation}}
\newcommand{\bea}{\begin{eqnarray}}
\newcommand{\eea}{\end{eqnarray}}
\newcommand{\eps}{\varepsilon}
\newcommand{\bk}{{\bf k}}
\newcommand{\bp}{{\bf p}}
\newcommand{\bq}{{\bf q}}

\newcommand{\al}{\alpha}
\newcommand{\bt}{\beta}
\newcommand{\ga}{\gamma}

\newcommand{\sg}{{\bm \sigma}}
\newcommand{\stau}{{\bm \tau}}
\newcommand{\pd}{\partial}
\newcommand{\Sp}{\mathrm{Tr}}

\newcommand{\dl}{\delta}
\newcommand{\Dl}{\Delta}
\newcommand{\om}{\omega}

\newcommand{\fl}{\mbox{\scriptsize FL}}

\newcommand{\lb}{\mbox{\scriptsize LB}_2}
\def\J{{\cal J}}

\begin{document}


\title{Different scenarios of topological phase transitions in homogeneous neutron matter}
\author{S.~S.~Pankratov}
\affiliation{National Research Centre Kurchatov Institute, pl. Akademika Kurchatova 1, Moscow, 123182, Russia}
\author{M.~Baldo}
\affiliation{Istituto Nazionale di Fisica Nucleare, Sezione di Catania, 64
Via S.-Sofia, I-95123 Catania, Italy}
\author{M.~V.~Zverev}
\affiliation{National Research Centre Kurchatov Institute, pl. Akademika Kurchatova 1, Moscow, 123182, Russia} \affiliation{Moscow Institute of Physics and
Technology, Institutskii per. 9, Dolgoprudnyi, Moscow region, 141700 Russia}

\date{\today}

\begin{abstract}
We study different scenarios of topological phase transitions in
the vicinity of the $\pi^0$ condensation point in neutron matter. The
transitions occur between the Fermi-liquid state and a
topologically different one with two sheets of the Fermi surface.
Two possibilities of a rearrangement of quasiparticle degrees of
freedom are shown: the first-order topological phase transition
and the second-order one. The order of the phase transition is
found to be strongly dependent on the value of the critical wave
vector of the soft $\pi^0$ mode. The thermodynamics of the system
is also studied. It is shown that the topology of the
quasiparticle momentum distribution is mainly determined by the
neutron matter density, while the temperature $T$ is essential in
a narrow density region. A simple explanation of the first-order
topological phase transition at $T=0$ is given.
\end{abstract}

\pacs{ 21.65.-f, 
26.60.-c, 
71.10.Ay 
}

\maketitle

\section{Introduction}

Dense neutron matter is an example of systems in which correlations of
single-particle degrees of freedom are strongly enhanced in a certain region
of external parameters. Such enhancement is attributed to an exchange of critical
fluctuations of a very soft collective mode in a vicinity of its
collapse \cite{Dugaev1976}. In dense neutron matter, critical
spin-isospin fluctuations with quantum numbers of the neutral pion are enhanced
close to the $\pi^0$ condensation point (PCP) \cite{MigdalRevModPhys1978}.
Critical density for the neutral pion condensation $\rho_c\simeq 0.2$\,fm$^{-3}$
predicted in microscopic calculations \cite{Wiringa,AkmalPandRav1998}
is reached in a typical neutron star with the central density of
$0.5-1.0$\,fm$^{-3}$.

Strong momentum dependence of the quasiparticle (QP) interaction due to exchange
of critical spin-isospin fluctuations may result in a change in the topology
of the ground state of a neutron QP system \cite{Voskr2000}. To our knowledge,
the possibility of a change in the topology of the ground state was discussed
for the first time in Ref.\ \cite{Frohlich-PR-1950} for electronic
systems. In nuclear physics, QP momentum distributions $n(\bp)$ with a topology
different from that of the Fermi sphere $n_{\fl}(\bp)=\theta(p_F-p)$ were
considered in Refs.\ \cite{Vary-PRC-1979,Zabolitsky-PRC-1979} for model
interactions and in Ref.\ \cite{Aguilera-PRC-1980} for semirealistic ones.

\begin{figure}[t]
\begin{minipage}[h]{0.49\linewidth}
\begin{center}
\includegraphics[width=0.95\linewidth,height=0.7\linewidth]{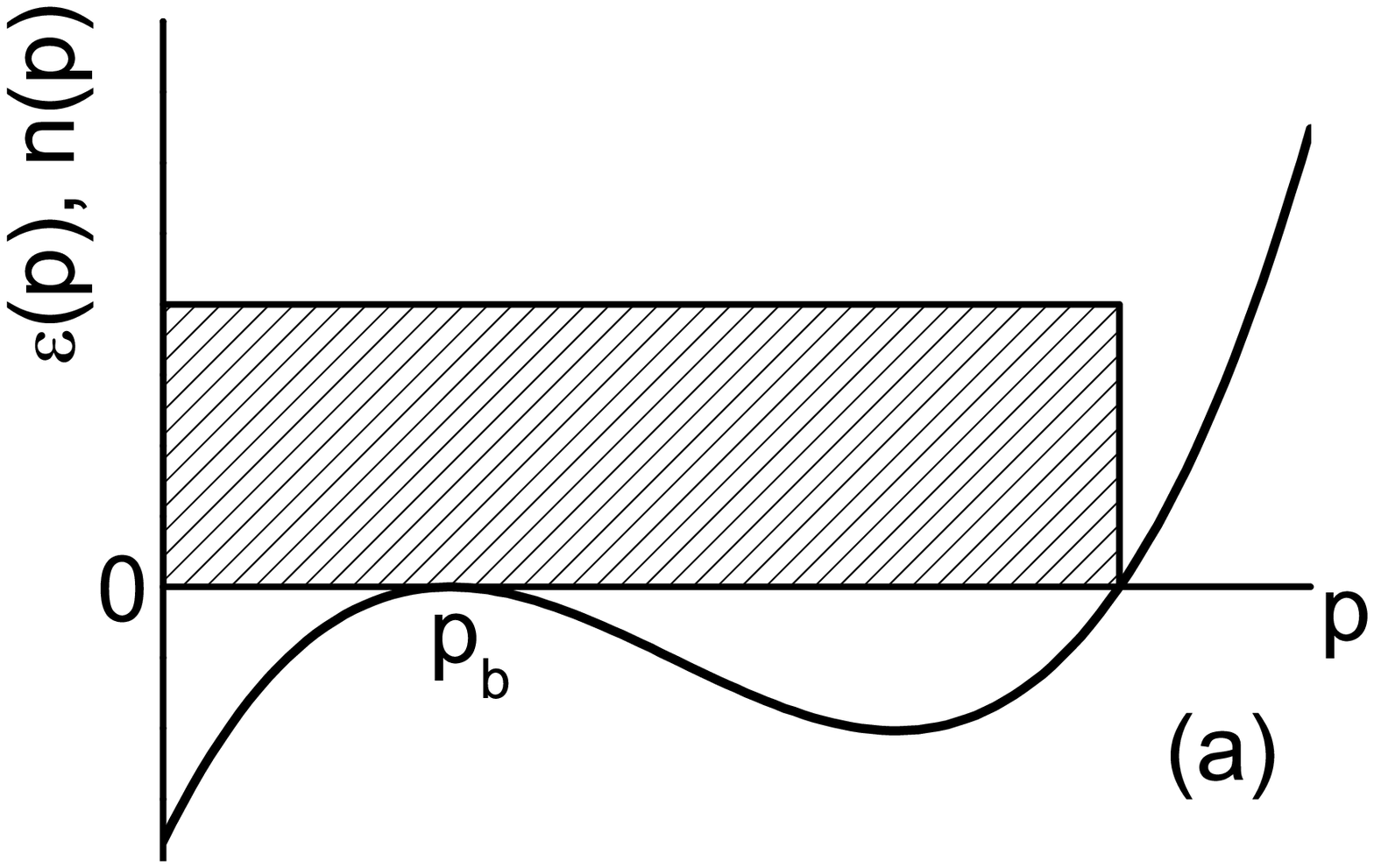}
\end{center}
\end{minipage}
\hfill
\begin{minipage}[h]{0.49\linewidth}
\begin{center}
\includegraphics[width=0.95\linewidth,height=0.7\linewidth]{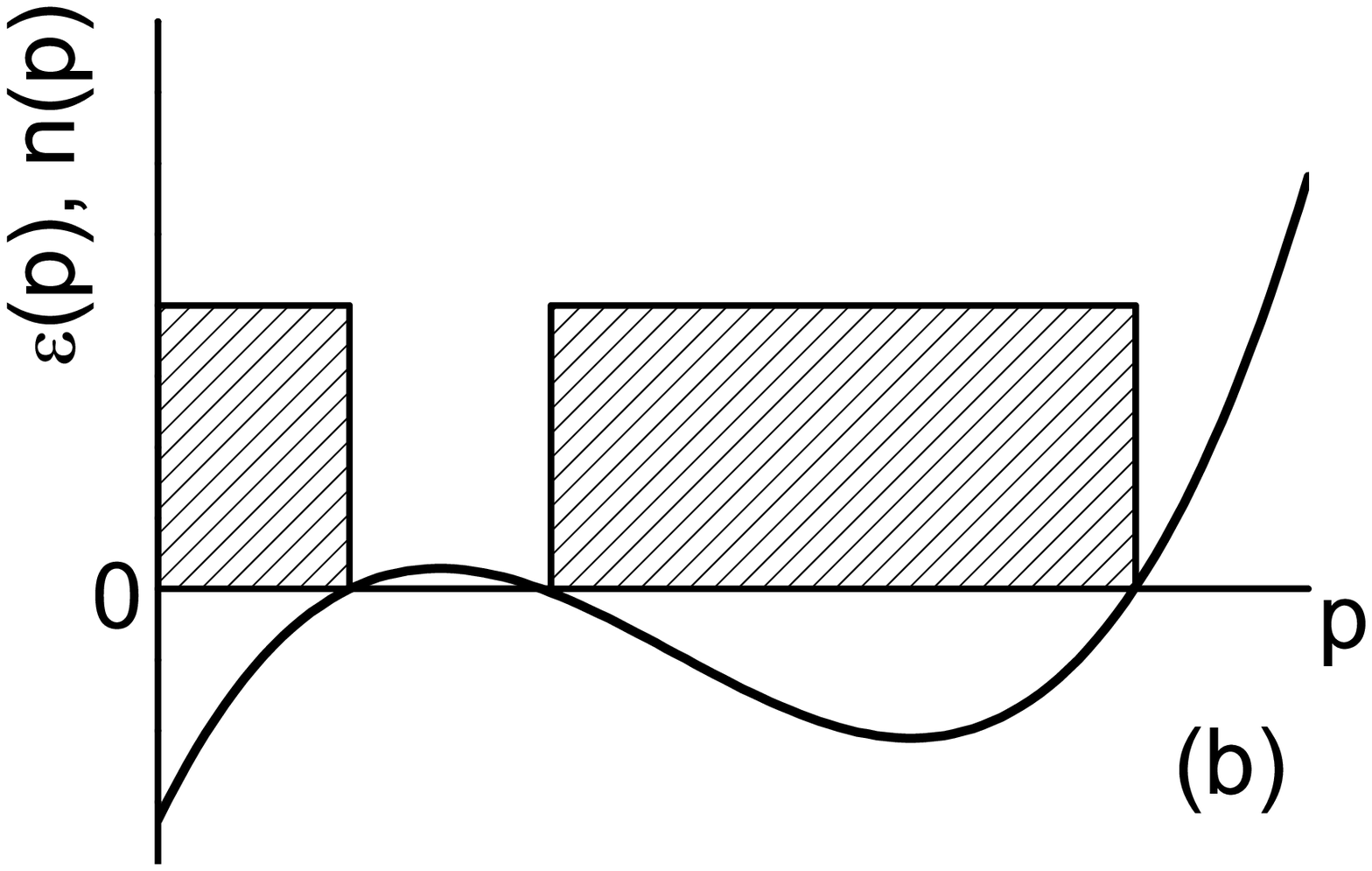}
\end{center}
\end{minipage}
\begin{center}
\begin{minipage}[c]{0.9\linewidth}
\caption{Left panel: QCP at the $\rho_b$ density value. Right panel: Beyond the QCP.} \label{fig:spqcp}
\end{minipage}
\end{center}
\end{figure}

The non-Fermi-liquid topology of the ground state QP momentum
distribution in the vicinity of PCP was first discussed in Ref.\ \cite{Voskr2000}. With the increase of the density $\rho$ towards
the PCP value $\rho_c$ the QP spectrum $\eps(\bp)$ (measured from
the chemical potential $\mu$) becomes a nonmonotonic function and
at certain density $\rho_b$ touches the momentum axis at
some point $p_b$ (see panel (a) of Fig.~\ref{fig:spqcp}). This
situation is associated with a quantum critical point (QCP)
\cite{KhodelTwoScen2007} at which the single-particle density of
states diverges \cite{KhodelTopCros2011}. Beyond the QCP, the
Landau state with the Fermi step QP distribution $n_{\fl}(\bp)$
becomes unstable as it violates the necessary stability condition
\beq \delta E[n]=2\int
\eps(\bp,[n(\bp)])\,\delta n(\bp)\,d\upsilon>0\,.\label{necessary_condnucl}
\eeq
Here $d\upsilon={d^3\bp}/{(2\pi)^3}$ is the volume element in the momentum
space and the factor 2 stands for summation over spin projections.
The constraint (\ref{necessary_condnucl}) requires a positive change in the total energy $E[n]$ of the system for any admissible variation $\delta n(\bp)$ of the QP momentum distribution that conserves the density \beq 2\int\delta n(\bp)\,d\upsilon=0\,.\eeq
Thus a new state appears with an unoccupied region (\lq\lq bubble\rq\rq) in the
momentum distribution, which has several sheets of the Fermi
surface (panel (b) of Fig.~\ref{fig:spqcp}). This is the state that was considered
in Refs.\ \cite{Vary-PRC-1979,Zabolitsky-PRC-1979,Aguilera-PRC-1980}.
Further development concerning states with many bubbles can be found
in Refs.\ \cite{ZverBaldo1998,ArtPogShag1998}. It is also worth noting investigations \cite{Aguilera-PRC-1982} concerning abnormal occupation
in boson matter.
%
%

Reconstructions of the QP momentum distribution in strongly
correlated Fermi systems changing the Fermi surface
topology are generally referred \cite{Volovik-book-2007} to as
topological phase transitions (TPTs). It should be noted that, besides the bubble scenario, there is another type of TPT which is called the fermion condensation \cite{KS-JETPL-1990,Volovik-JETPL-1991,Nozieres-1992}.
In this scenario, the QP spectrum acquires a flat band $\eps(p)=0,\; p\in[p_i,p_f]$,
and the Fermi surface changes its dimension. The relation between
the bubble scenario and the fermion condensation is discussed in Ref.\ \cite{KhodelQCP2008}.

Going back to the neutron matter problem, we note that a bubble
formation beyond the QCP is quite important for cooling of
neutron stars \cite{Voskr2000,VAKhod2004}. Indeed, a new sheet
of the neutron Fermi surface with a low value of the corresponding Fermi momentum plays an
important role for operation of the direct Urca processes:
$n\rightarrow p+e^{-}+\bar{\nu_e},\; p+e^{-}\rightarrow n+\nu_e$.
In a neutron star's core these processes are generally considered
to be forbidden \cite{Yk-book-2007} due to the kinematic
restriction on the Fermi momenta of the involved particles,
$p_{F_n}\le p_{F_p}+p_{F_e}$. In the typical density range of the
core $\sim 1-2\,\rho_0$ ($\rho_0\simeq0.16$\,fm$^{-3}$ is the
normal nuclear density), the proton fraction does not exceed $6-8\%$,
and the right-hand side of the kinematic equality is estimated by
$\sim 0.8\,p_{F_n}$. The appearance of the new sheet of the Fermi
surface at the point $p^{(1)}_{F_n}<0.8\,p_{F_n}$ provides agreement with
the kinematic restriction. This mechanism was considered in Refs.\ \cite{Voskr2000,VAKhod2004} as a possibility for the enhanced cooling of some neutron stars (e.g., Vela, Geminga, and 3C58).

\section{Quasiparticle approach near pion condensation point}

\subsection{General relations}

The method we use for a description of neutron matter near the PCP is based
on an implementation of the Landau-Migdal QP approach to strongly correlated
Fermi systems that is reviewed in details in Ref.~\cite{Khod2011MigAnn}. Within this
approach, the QP spectrum and the QP momentum distribution at finite temperature $T$
are evaluated by solving the set of equations,
\bea
&&{\partial\eps(\bp)\over\partial\bp} = {\bp\over
m} + \int\!f(\bp,\bp')\,{\partial n(\bp')\over\partial\bp'}\,
d\upsilon'\,,\label{lansp}\\
&&n(\bp)=\left[1+e^{\eps(\bp)/T}\right]^{-1}\,,
\label{eqn(p,T)}\\
&&2\int n(\bp)\, d\upsilon = \rho\,.
\label{nuclnorm}
\eea
The first equation of this set is the Landau relation where $f(\bp,\bp')$ is
the QP interaction function \cite{Landau,Abrikos} and $m$ stands for the free
neutron mass; the second equation is the Fermi-Dirac formula in which
$\eps(\bp)$ is a functional of $n(\bp)$; and the last one is the normalization
condition.

\begin{figure}[t]
\begin{center}
\includegraphics[width=0.95\linewidth,height=0.24\linewidth]{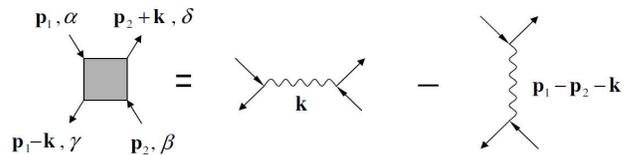}
\begin{minipage}[c]{0.9\linewidth}
\caption{The main contribution to the QP interaction function.} \label{fig:Dug}
\end{minipage}
\end{center}
\end{figure}

The QP interaction function is identified \cite{Landau,Abrikos} with the
$\om$ limit of the vertex function $\Gamma$,
\begin{multline}
f(\bp_1,\sg_1;\bp_2,\sg_2)=Z^2\Gamma^{\om}_{\al\bt,\ga\dl}(\bp_1,\bp_2)
\\=Z^2\lim_{\frac{k}{\om}\rightarrow0,\om\rightarrow0}
\Gamma_{\al\bt,\ga\dl}(\bp_1,\bp_2,\bk,\om)\,,
\end{multline}
where $Z$ is the residue of the single-particle Green function and
$\sg_{\al\ga}$ stands for Pauli spin matrices. According to Ref.\ \cite{Dugaev1976}, the most singular contribution to the vertex function near the PCP comes from an exchange of a soft spin-isospin collective mode. The corresponding direct and the exchange graphs
are shown in Fig.~\ref{fig:Dug}. At the limit $k\rightarrow0$, the contribution
of the exchange graph still exhibits a strong dependence on the relative
momentum $\bq=\bp_1-\bp_2$. Therefore the QP interaction
function reads \cite{Voskr2000}
\beq
f(\bp_1,\sg_1;\bp_2,\sg_2)\simeq(\J_{\al\dl}
D\J_{\bt\ga})(\bq,\om=0;\rho)\label{dlFsing}\,,
\eeq where
$\J_{\al\dl}$ is the interaction vertex of nucleons and pions in
neutron media, $D$ is the $\pi^0$ propagator, and the general
arguments of the operators are in parentheses.
%
%

\subsection{Topological phase transitions}

The occurrence of TPTs in dense neutron matter can be traced with the help
of the strong momentum dependence of the QP interaction function near
the PCP. As was discussed in Ref.\ \cite{MigdalRevModPhys1978}, the spectrum
$\om(q)$ of $\pi^0$-like collective excitations in neutron matter is given
by a particular branch of poles of the $\pi^0$ propagator. The behavior of
this branch depends on the density $\rho$ of the system. Pion condensation
occurs at the critical density $\rho_c$ where the excitation energy vanishes,
$\om(q_c)=0$, together with its derivative, $\pd\om(q_c)/\pd q=0$,
at a certain momentum $q_c$. As a consequence, the following conditions
at the PCP are valid:
\beq
D^{-1}(q_c,0;\rho_c)=0\,,\quad \left.\dfrac{\pd D^{-1}(q,0;\rho_c)}{\pd q^2}\right|_{q_c}=0\,.\eeq Thus using the Taylor expansion of $D^{-1}$, the interaction function $f$ entering in Eq.~(\ref{lansp}) can be written \cite{Voskr2000,VAKhod2004,Baldo2004} in the form \begin{multline} f(\bp_1,\bp_2)=\frac{1}{2}\,\Sp_{\sg_1}\Sp_{\sg_2}f(\bp_1,\sg_1;\bp_2,\sg_2)
\\=\dfrac{g}{\kappa^2(\rho)+\left((\bp_1-\bp_2)^2/q_c^2-1\right)^2}\,,\label{fampl}\end{multline} where $g$ is an effective coupling constant and $\kappa^2(\rho)\propto(\rho_c-\rho)$ is an effective radius in momentum space. The notation $\Sp_{\sg}$ stands for the trace over the spin variable.

\begin{figure}[t]
\begin{center}
\includegraphics[width=0.75\linewidth,height=0.6\linewidth]{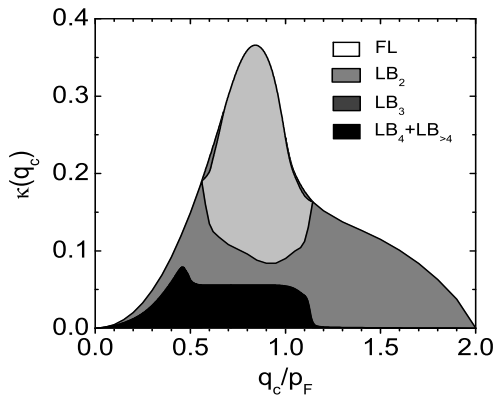}
\begin{minipage}[c]{0.9\linewidth}
\caption{The phase diagram of the neutron matter ground state near the PCP \cite{Baldo2004}.}\label{fig:Baldo2004}
\end{minipage}
\end{center}
\end{figure}

Previous investigations \cite{Baldo2004} within the QP interaction (\ref{fampl}) were
focused on the ground-state topology, and $g,q_c,\kappa$
quantities were regarded as phenomenological parameters.
For convenience of readers, we present in Fig.~\ref{fig:Baldo2004}
a topological phase diagram in $q_c,\kappa$ variables obtained in that work.
The label FL corresponds to the Fermi-liquid state and LB$_i$, to states
with $i$ sheets of the Fermi surface.
%
%
A change in the density $\rho$ leads to a change in the system
position $(q_c,\kappa)$ on the diagram. Transitions between
different regions of the phase diagram represent TPTs that can
occur in neutron matter. Until recently, all such TPTs were
considered \cite{KhodelQCP2008,Khod2011MigAnn} to occur
continuously according to violation of
Eq.~(\ref{necessary_condnucl}). However, an attentive investigation
\cite{PanZver2012} within the model (\ref{fampl}) revealed another possibility, namely,
a first-order TPT. Such a scenario of the Fermi surface reconstruction in a
homogeneous isotropic Fermi system was first found
\cite{PanZverBaldo2011} in a model of strongly correlated 2D
electron gas beyond the QCP. Below, we present a detailed analysis
of possible TPTs for the neutron matter problem.

\section{Semi-microscopic QP interaction function}

The above discussion of TPTs was restricted by the phenomenological
description (\ref{fampl}) of the QP interaction function. It is
possible to convey neutron matter physics near the PCP in a more
direct way. A semi-microscopical expression for the QP interaction
can be derived by use of the microscopic formula (\ref{dlFsing}).
%
\begin{figure}[t]
\begin{center}
\includegraphics[width=0.8\linewidth,height=0.6\linewidth]{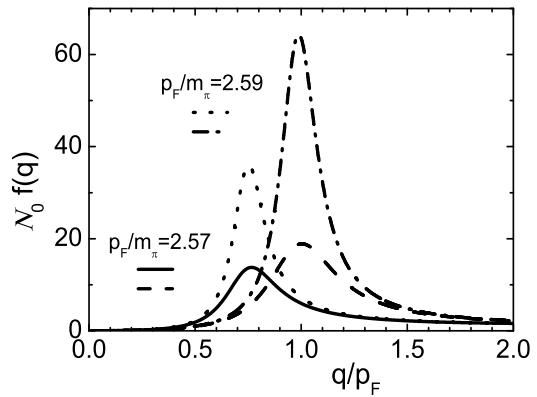}
\begin{minipage}[c]{0.9\linewidth}
\caption{The interaction function (\ref{micrampl}) multiplied by
${\cal N}_0=mp_F/\pi^2$ as a function of $q/p_F$. The parametrization (\ref{PMigParam}) corresponds to the curves peaked at $q_c\simeq0.74\,p_F$, while (\ref{PthisParam}) corresponds to $q_c\simeq p_F$.}
\label{fig::ampl}
\end{minipage}
\end{center}
\end{figure}
%

The bare $\pi^0$-nucleon interaction vertex is given
\cite{PionBook} by \beq J^{\,0}_{\pi NN}=\dfrac{if}{m_\pi}({\bm
\sigma}\bq)\tau_{3}\,,\eeq where $f\simeq1$ is the dimensionless
$\pi^0$-nucleon coupling constant, $m_\pi$ is the pion mass, and
$\tau_{3}$ is the diagonal isospin matrix. The vertex
renormalization in neutron matter is due to nucleon-nucleon
correlations that can be described by means of the Landau-Migdal
amplitude \cite{MigdalTKFS}
\beq\dfrac{m^*p_0}{\pi^2}\mathcal{F}=F+F'{\bm \tau_1}{\bm
\tau_2}+(G+G'\stau_1\stau_2)\sg_1\sg_2\,.\eeq Here
$p_0=\left(1.5\pi^2 \rho_0\right)^{1/3}$ is the Fermi momentum in
equilibrium nuclear matter and $m^*$ is the effective nucleon
mass. According to Ref.\ \cite{MigdalRevModPhys1978}, the renormalized
static vertex reads\beq
J^{st}_{\al\bt}(\bq)=\dfrac{if}{m_\pi}\dfrac{({\bm
\sigma}_{\al\bt}\bq)\tau_{3}}{1+g_{c}\chi_{NN}^{st}(q)}\,,\quad
g_{c}=\dfrac{\pi^2}{m^*p_0}\dfrac{m_\pi^2}{f^2}(G+G')\,.\label{Jren}\eeq
The function $\chi_{NN}^{st}(q)$ is the static susceptibility of
free QPs proportional to the Lindhard function:
\begin{multline}\chi_{NN}^{st}(q)=\dfrac{f^2}{m_{\pi}^2}\dfrac{m^*p_F}{\pi^2}\\\times\left(\dfrac{1}{2}+\dfrac{p_F}{2q}\left(\dfrac{q^2}{4p_F^2}-1\right)\ln\left|\dfrac{1-q/2p_F}{1+q/2p_F}\right|\right)\,,\end{multline}
where $p_F=\left(3\pi^2 \rho\right)^{1/3}$ is the neutron Fermi
momentum.

The pion propagator in Eq.~(\ref{dlFsing}) is connected with the polarization
operator: $D^{-1}(q,\om)=\om^2-\bq^2-m_{\pi}^2-\Pi(q,\om)$. The
microscopic description of the pion polarization operator is a
quite subtle matter \cite{PionBook}. We use here its
semi-microscopic representation \cite{MigdalMarkMish1974}
\beq\Pi(q,0)=-q^2\left(\dfrac{\chi_{NN}^{st}(q)}{1+g_c\chi_{NN}^{st}(q)}+
\dfrac{\rho}{\rho_{\Dl}\left(1+q^2/q_{\Dl}^2 \right)}\right)\,.
\label{PolarOp}\eeq
The first term describes processes of particle-hole excitations
where nucleon-nucleon correlations are taken into account by the
denominator. The second term is a phenomenological one and
describes $\Delta$-resonance-hole excitations. The $S$-scattering processes in neutron matter are neglected.

Finally, one arrives at the formula \begin{multline} f(q)=\frac{1}{2}\,\Sp_{\sg_1}\Sp_{\sg_2}f(\bp_1,\sg_1;\bp_2,\sg_2)\\ 
=\left(\dfrac{f}{m_\pi}\dfrac{q}{1+g_{c}\chi_{NN}^{st}(q)}\right)^2\,\dfrac{1}{m_{\pi}^2
+ q^2 + \Pi(q,0)}\,\label{micrampl}\end{multline} for the interaction function, where the polarization operator is given by (\ref{PolarOp}). In Eq.~(\ref{micrampl}), the constants $f=1$, $m^{*}=m$, $G+G'=1$ are fixed all along the
further discussion, while two sets of parameters for
the phenomenological part of the polarization operator are used. The first
set,
\beq \rho_{\Dl}=0.59\,m_{\pi}^3\,,\quad
q_{\Dl}=2.08\,m_{\pi}\,,\label{PMigParam}\eeq
corresponds to Ref.~\cite{MigdalMarkMish1974}, while the second one,
\beq
\rho_{\Dl}=0.97\,m_{\pi}^3\,,\quad
q_{\Dl}=4.1\,m_{\pi}\,,\label{PthisParam}\eeq
is suggested in this work.
Both sets reproduce the value $\rho_c\simeq0.2$\,fm$^{-3}$ that agrees with
the critical density of $\pi^0$ condensation obtained in
\cite{Wiringa,AkmalPandRav1998}. The difference between the two parametrizations
is in the corresponding value of the critical wave vector $q_c$, which is
not known accurately from microscopic calculations. We also note that
$\rho_c$ and $q_c$ are quite sensitive to tuning of
parameters $\rho_{\Dl}$ and $q_{\Dl}$. The behavior of the QP
interaction function is shown in Fig.~\ref{fig::ampl}. The
parametrization (\ref{PMigParam}) leads to $q_c\simeq0.74\,p_F$
(that is less then $p_F$), while (\ref{PthisParam}) leads to $q_c\simeq
p_F$. Figure \ref{fig::ampl} also demonstrates the amplification of
the QP interaction with an increase of $p_F$ (equally, the density
$\rho=p_F^3/3\pi^2$). The divergence is reached at the PCP point, $p_{Fc}\simeq2.602\,m_{\pi}$ and
$p_{Fc}\simeq2.598\,m_{\pi}$ correspondingly for the first and the
second parametrizations.

\section{Different scenarios of topological phase transitions}

In this Section, we discuss results of the analysis of topological rearrangements of QP degrees
of freedom based on the semi-microscopic QP interaction function. The QP spectrum and the QP momentum distribution are determined by the set of Eqs.\ (\ref{lansp}),(\ref{eqn(p,T)}),(\ref{nuclnorm}). Due to the assumed dependence of the interaction function (\ref{micrampl}) on the relative momentum $q$, Eq.~(\ref{lansp}) can be integrated by parts, yielding \beq \eps(\bp) = {p^2\over 2m} -\mu + \int\! f(\bp-\bp')\,n(\bp')\,d\upsilon'\,.\label{eqeps}\eeq For solving the equations a contracting iterations method was used. The case of $T=0$ was modeled by $T=10^{-5}\eps_F^0$, where $\eps_F^0=p_F^2/2m$.

\begin{figure}[t]
\begin{minipage}[h]{0.49\linewidth}
\begin{center}
\includegraphics[width=1\linewidth,height=0.9\linewidth]{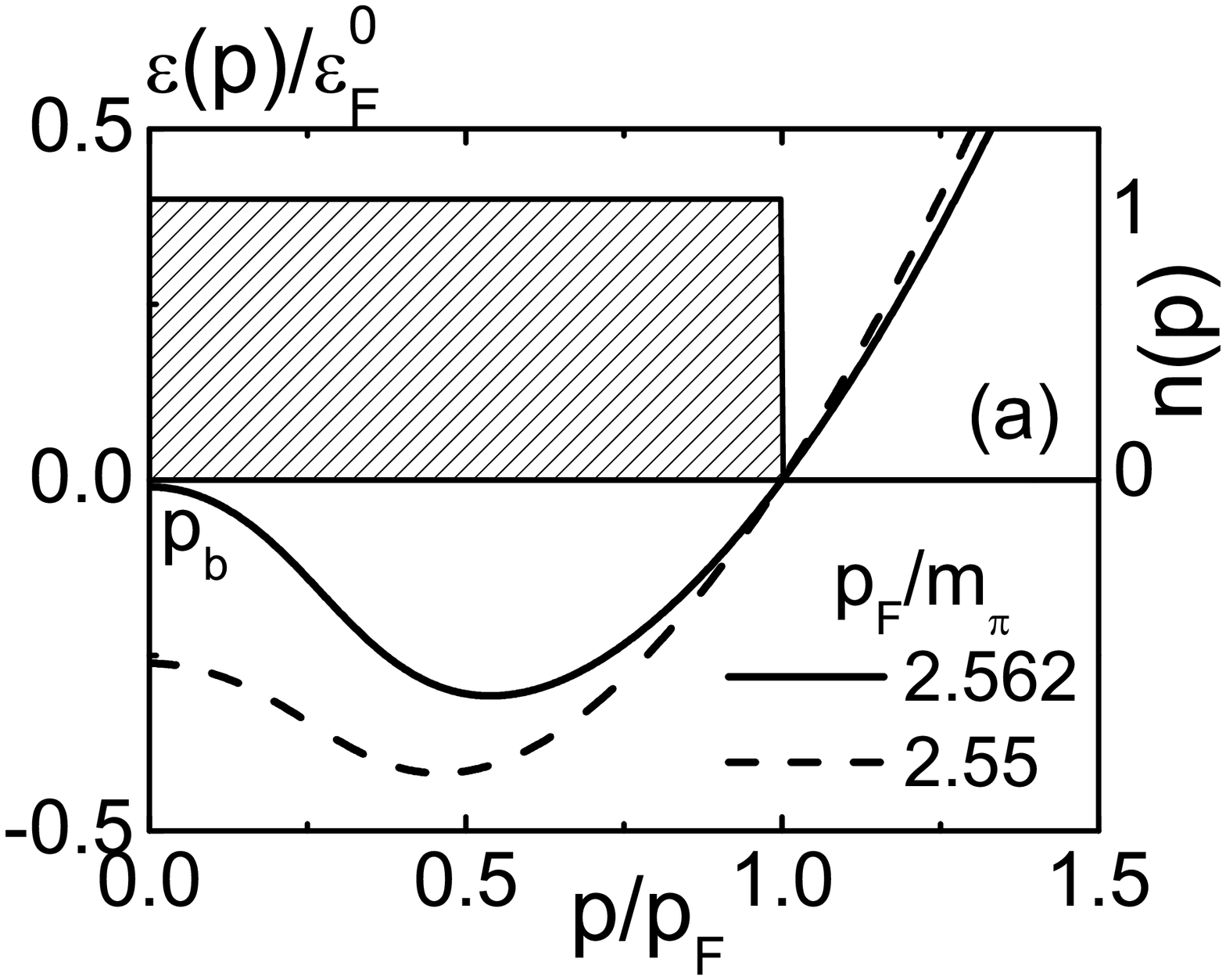}
\end{center}
\end{minipage}
\hfill
\begin{minipage}[h]{0.49\linewidth}
\begin{center}
\includegraphics[width=1\linewidth,height=0.9\linewidth]{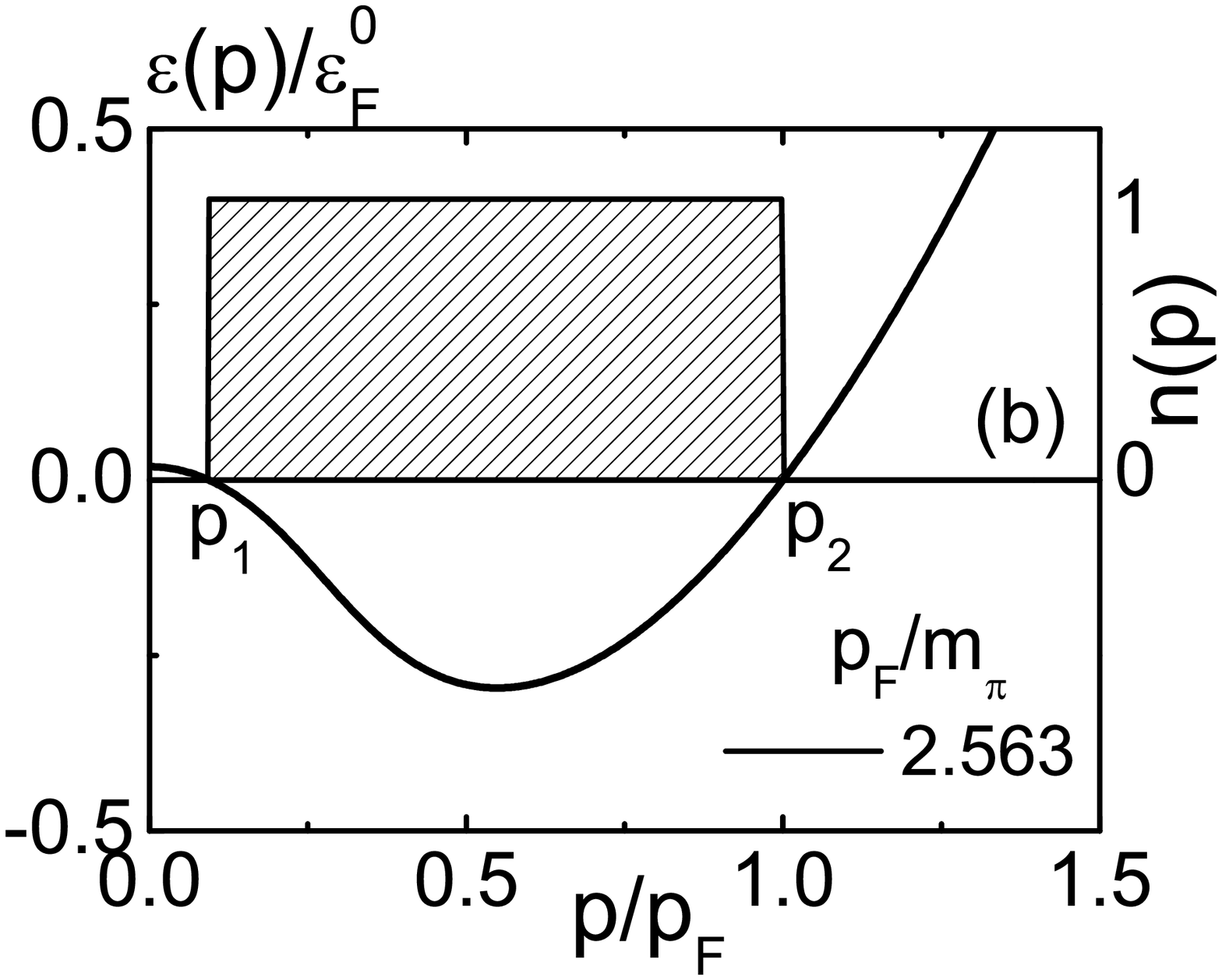}
\end{center}
\end{minipage}
\begin{center}
\begin{minipage}[c]{0.9\linewidth}
\caption{Evolution of the QP spectrum (in units $\eps^0_F=p_F^2/2m$) and the QP momentum distribution with increasing $p_F$ within the model (\ref{PMigParam}). The temperature $T=0$.}\label{fig:model1spectT0}
\end{minipage}
\end{center}
\end{figure}

First, we consider the results obtained within the model (\ref{PMigParam}) where $q_c=0.74\,p_F$. The evolution of QP degrees of freedom, driven by an increase of $p_F$, is displayed in Fig.~\ref{fig:model1spectT0}.
%
%
It has a continuous behavior in agreement with the scenario of TPTs with violation
of the necessary stability condition (\ref{necessary_condnucl}). As the Fermi momentum reaches the critical value $p_F^b\simeq2.562\,m_{\pi}$, a bifurcation occurs, and a new zero of the
QP spectrum $\eps(p)$ appears at the momentum $p_b=0$. Beyond the $p_F^b$, the QP
momentum distribution possesses two sheets of the Fermi surface
with coordinates $p_1\ll p_F$ and $p_2\simeq p_F$. The size of the bubble region
(which is equal to $p_1$) increases continuously from the zero
value with further increase of $p_F$.

%
\begin{figure}[]
\begin{minipage}[h]{0.49\linewidth}
\begin{center}
\includegraphics[width=1\linewidth,height=0.9\linewidth]{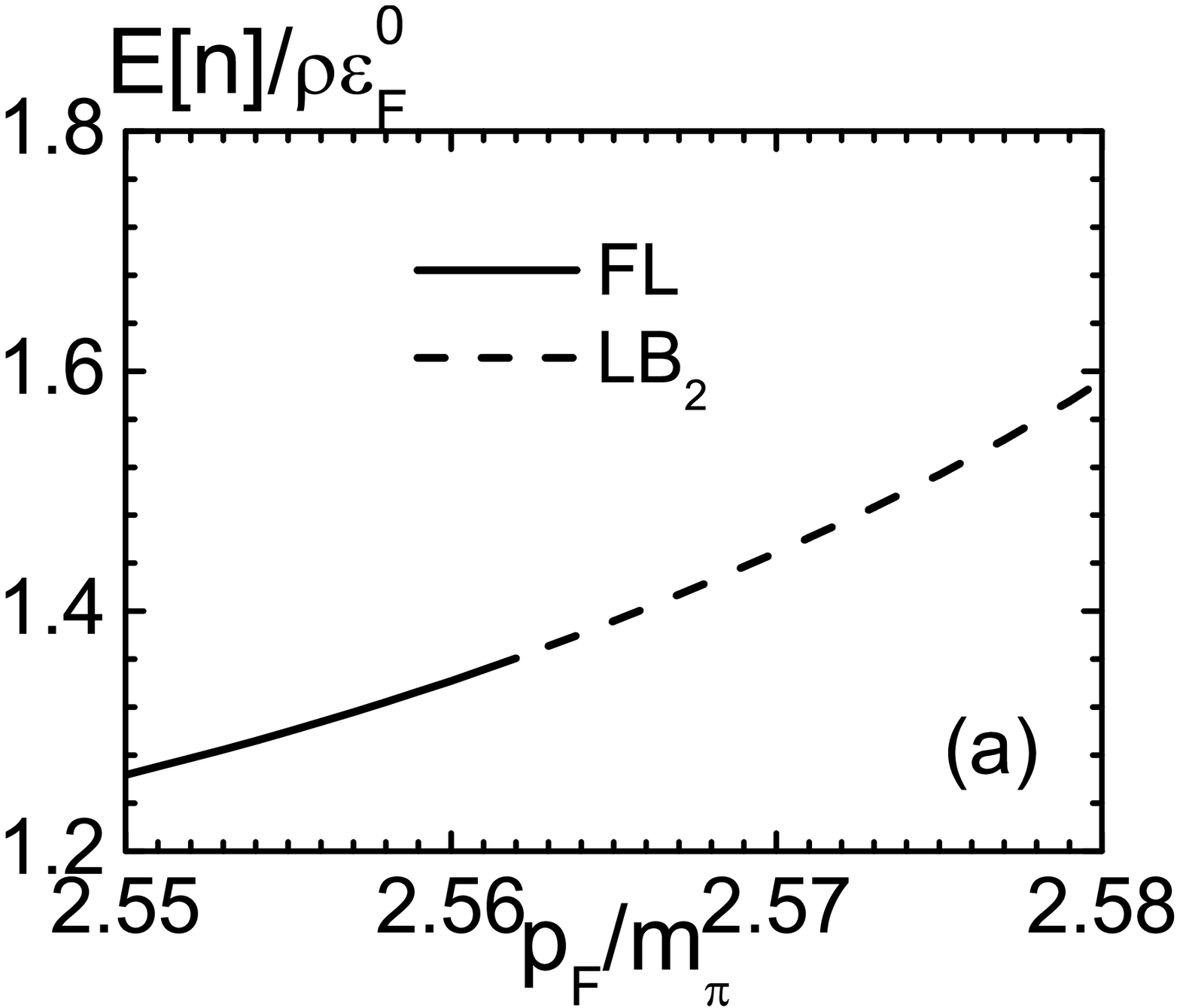}
\end{center}
\end{minipage}
\hfill
\begin{minipage}[h]{0.49\linewidth}
\begin{center}
\includegraphics[width=1\linewidth,height=0.9\linewidth]{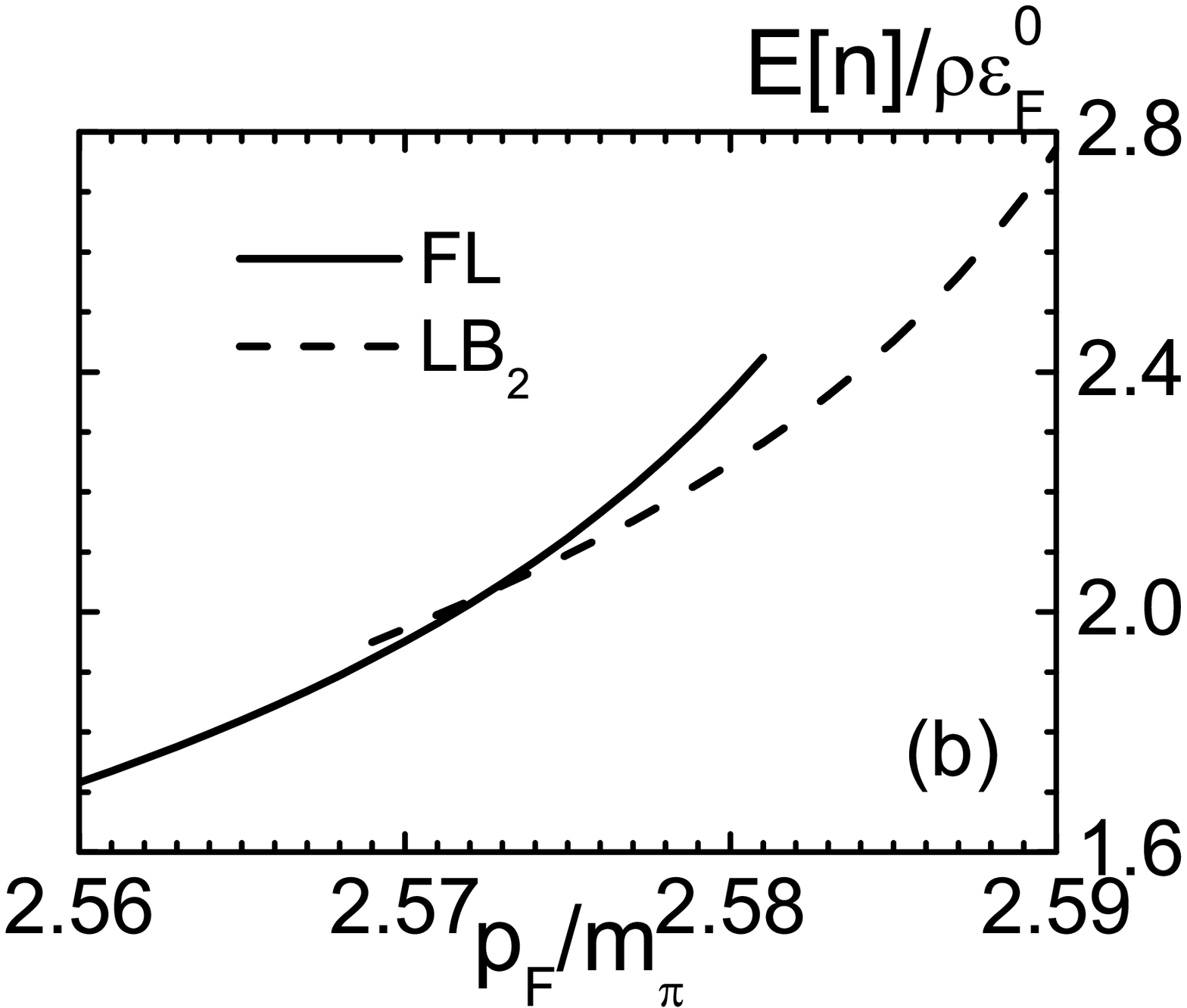}
\end{center}
\end{minipage}
\begin{center}
\begin{minipage}[c]{0.9\linewidth}
\caption{Energy per particle at $T=0$ as a function of $p_F$. The left panel corresponds to the model (\ref{PMigParam}), while right one corresponds to the model (\ref{PthisParam}).}\label{fig:EnrT0}
\end{minipage}
\end{center}
\end{figure}
%
%
\begin{figure}[]
\begin{center}
\includegraphics[width=1\linewidth,height=0.8\linewidth]{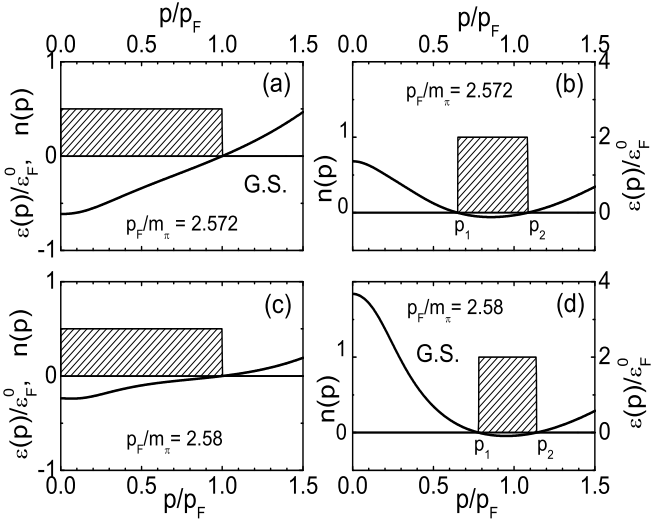}
\begin{minipage}[c]{0.9\linewidth}
\caption{Panels (a) and (b): QP spectra and QP momentum distributions of coexisting states
in advance of the first-order TPT. Panels (c) and (d): The same beyond the TPT point. The temperature $T=0$.}
\label{fig::model2spectT0}
\end{minipage}
\end{center}
\end{figure}
%
Dealing with Eq.~(\ref{eqeps}) for the QP spectrum, one can consider the
corresponding energy functional \beq E[n] =
2\int\dfrac{p^2}{2m}\,n(\bp)\,d\upsilon + \int
f(\bp-\bp')\,n(\bp)\,n(\bp')\,d\upsilon\,d\upsilon'\,.\label{nuclenergy}\eeq
The behavior of the energy is given in panel (a) of Fig.~\ref{fig:EnrT0}. It shows
a monotonic increase according to the continuous picture of the evolution of the ground state
in Fig.~\ref{fig:model1spectT0}. A second-order topological transition occurs between the FL
and the LB$_2$ states exactly at the critical point $p_F^b\simeq2.562\,m_{\pi}$.

Passing to the model (\ref{PthisParam}) where $q_c\simeq p_F$, we first consider
panel (b) of Fig.~\ref{fig:EnrT0}. As it
is seen, there are two different states in the interval
$2.57\,m_{\pi}\lesssim p_F\lesssim2.58\,m_{\pi}$. The coexistence of several solutions of Eqs.\ (\ref{eqn(p,T)}),(\ref{nuclnorm}),(\ref{eqeps}) was established by considering different initial conditions for the iteration procedure. Fig.~\ref{fig::model2spectT0} represents QP spectra and QP momentum distributions of coexisting states. At the value
$p_F=2.572\,m_{\pi}$ the ground state is the FL one (panel (a)),
while there is also a LB$_2$ state (panel (b)) with a slightly
higher energy value. It is worth emphasizing that the bubble region has finite, not
negligible size $p_1\simeq0.6\,p_F$. A first-order phase transition occurs at $p_F^{1st}\simeq2.573\,m_{\pi}$. Beyond the transition point, the LB$_2$ state (panel (d)) becomes energetically favored over the FL state (panel (c)). Finally, the local energy minimum of the functional (\ref{nuclenergy}), corresponding to the FL state, becomes unstable \cite{footnote} (for $p_F\gtrsim2.581\,m_{\pi}$) and only the LB$_2$ ground state remains (see panel (b) of Fig.~\ref{fig:EnrT0}).

\begin{figure}[t]
\begin{center}
\includegraphics[width=1\linewidth,height=0.8\linewidth]{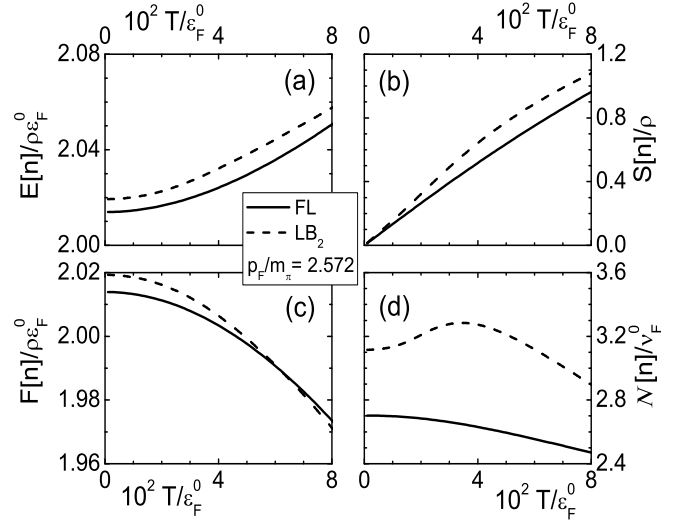}
\begin{minipage}[c]{0.9\linewidth}
\caption{The energy (a), the entropy (b), the free energy (c) and the density of states (d) as functions of the temperature for the FL and the LB$_2$ states.}
\label{fig::model2TermoF}
\end{minipage}
\end{center}
\end{figure}

We proceed with a thermodynamic analysis concerning the case of the first-order TPT.
Thermodynamic functions of neutron matter with the Fermi momentum
$p_F=2.572\,m_{\pi}$ are given by Fig.~\ref{fig::model2TermoF}.
%
%
Panel (a) demonstrates that the FL state is lower in the
energy than the LB$_2$ one up to the temperature
$T\sim0.1\,\eps_F^0\sim5\,$MeV. At the same time, panel (b) shows
that the entropy \beq S[n]=-2\int n(\bp)\ln n(\bp)+(1-n(\bp))\ln(1-n(\bp))\,d\upsilon\,,\eeq of
the LB$_2$ state grows more rapidly with increasing temperature than the entropy of the FL state. An interplay of contributions to the free energy $F[n] = E[n] - TS[n]$ leads to a first-order TPT driven by the temperature. Behavior of the free energy given by panel (c) demonstrates the first-order phase transition at the temperature $T_1\simeq6.2\cdot10^{-2}\eps_F^0$. Finally, panel (d) shows the temperature evolution of the density of states \beq{\cal N}[n] = \int\dfrac{d n(\bp)}{d\eps}\,d\upsilon=\dfrac{1}{T}\int n(\bp)(1-n(\bp))\,d\upsilon\,.\eeq

\begin{figure}[]
\begin{minipage}[h]{0.49\linewidth}
\begin{center}
\includegraphics[width=1\linewidth,height=0.9\linewidth]{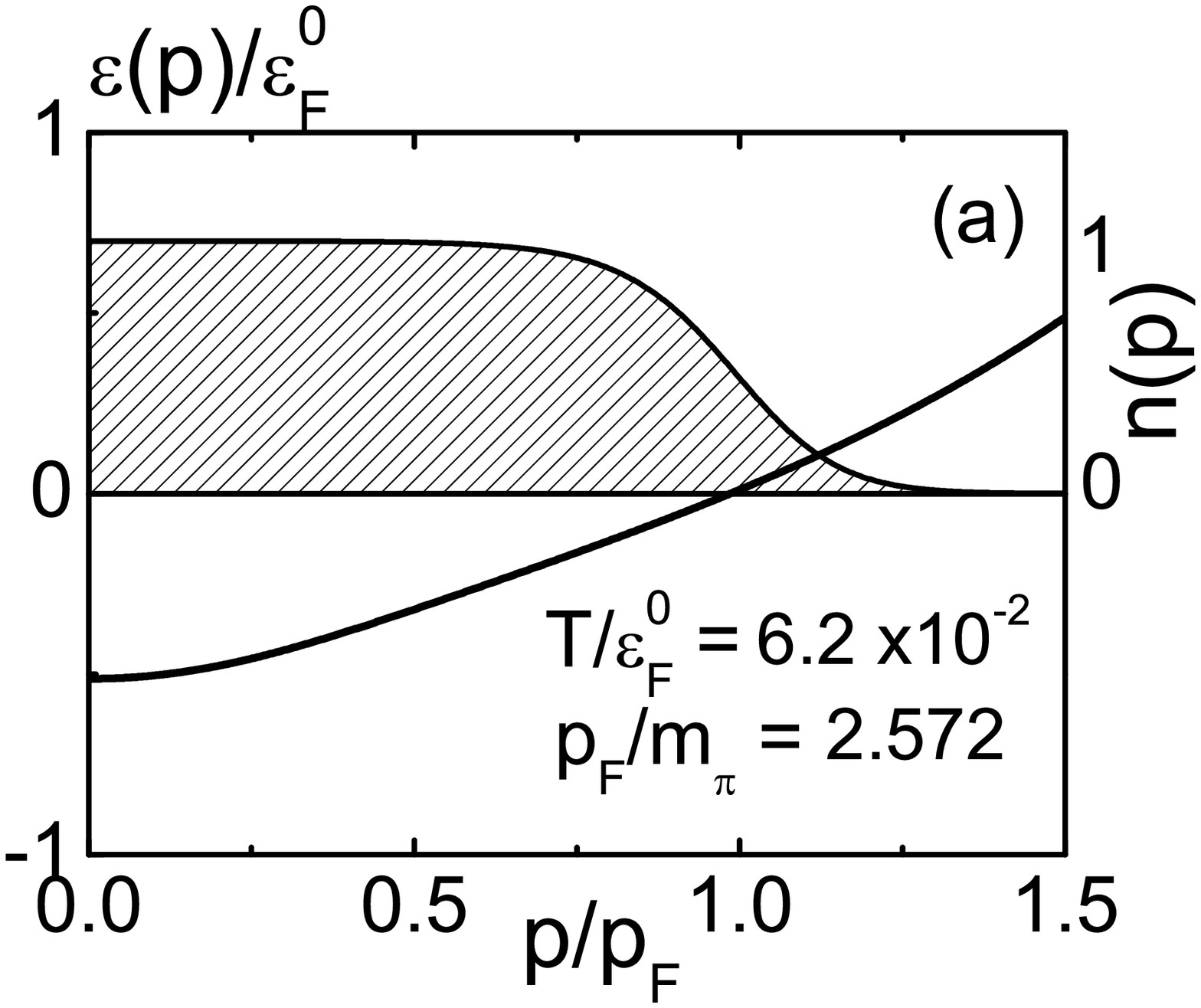}
\end{center}
\end{minipage}
\hfill
\begin{minipage}[h]{0.49\linewidth}
\begin{center}
\includegraphics[width=1\linewidth,height=0.9\linewidth]{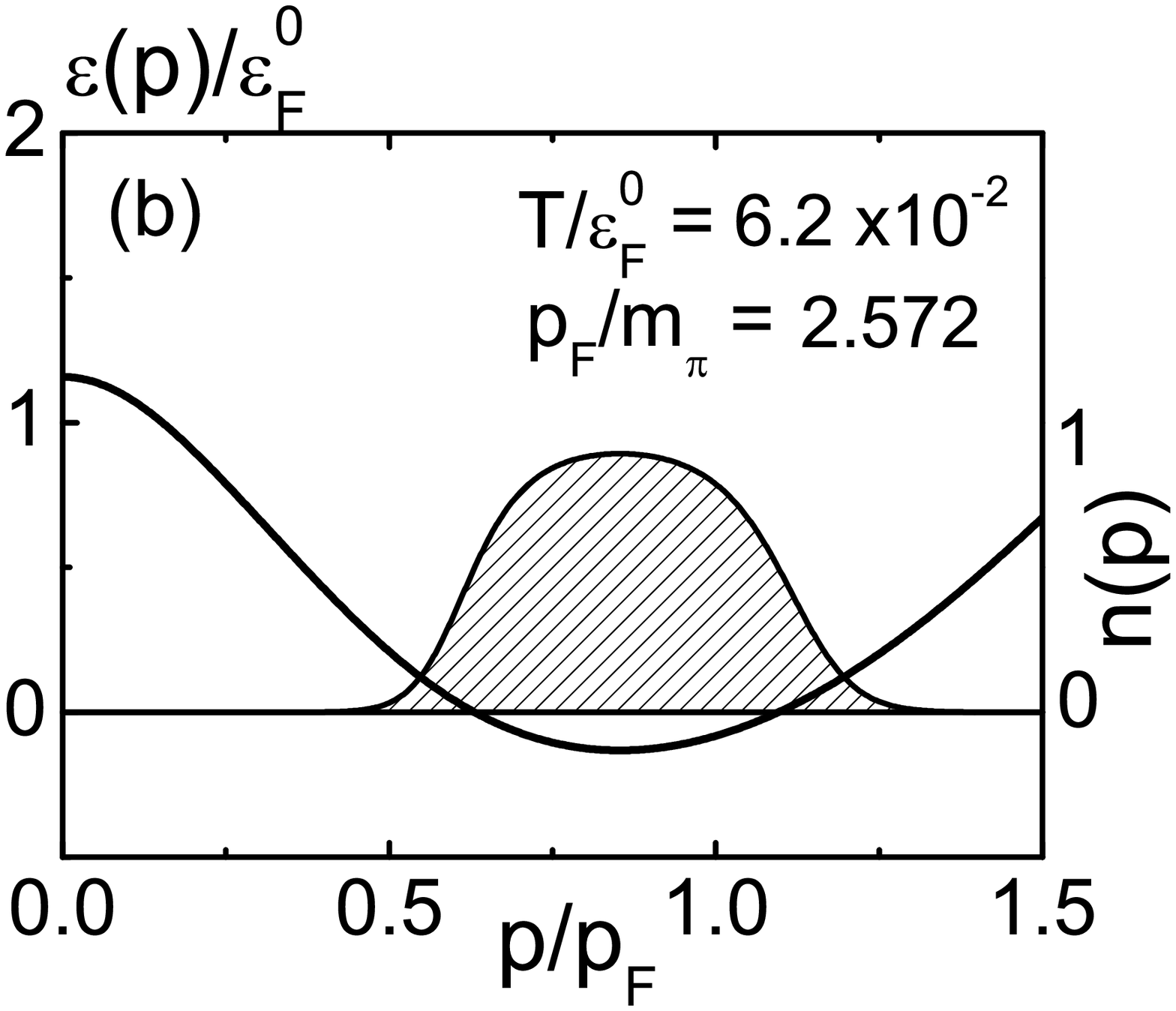}
\end{center}
\end{minipage}
\begin{center}
\begin{minipage}[c]{0.9\linewidth}
\caption{QP spectra and momentum distributions of the FL and the LB$_2$ states at the point (the black dot A in Fig.~\ref{fig:map} below) of the temperature-driven TPT.}\label{fig:model2T}
\end{minipage}
\end{center}
\end{figure}

To elucidate the temperature behavior of the thermodynamic functions under consideration, we present in Fig.~\ref{fig:model2T} QP spectra and QP momentum distributions of
the FL state and of the LB$_2$ one at the point $T=6.2\cdot10^{-2}\eps_F^0$ near the phase transition. This figure shows that the entropy and the density of states of the LB$_2$ state are larger than those of the FL state due to the fact that in the first case, two sheets of the Fermi surface \lq\lq melt\rq\rq, while in the second case only one does.

%
%
%
\begin{figure}[t]
\begin{center}
\includegraphics[width=0.8\linewidth,height=0.6\linewidth]{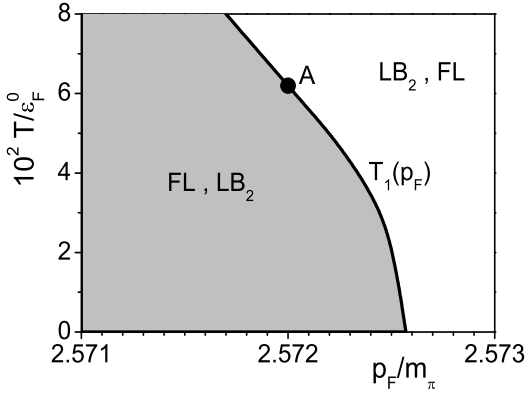}
\begin{minipage}[c]{0.9\linewidth}
\caption{Phase diagram of neutron matter near the line $T_1(p_F)$ of the fist-order TPTs. The first label corresponds to the thermodynamically favored phase, while the second one corresponds to a metastable phase. The black dot A refers to the QP states shown in Fig.~\ref{fig:model2T}.}\label{fig:map}
\end{minipage}
\end{center}
\end{figure}
%

The results of the analysis on the temperature behavior of QP states, in the
region near the point $p_F=2.572\,m_{\pi}$ (see panel (b) of
Fig.~\ref{fig:EnrT0}), are summarized by the $(p_F,T)$-phase diagram
of neutron matter shown in Fig.~\ref{fig:map}. The diagram
consists of two regions: the first one corresponds to the FL phase
of the system, while the other one to the LB$_2$ phase. The
regions are separated by the line $T_1(p_F)$ of the first-order
TPTs. The diagram shows that neutron matter state topology is
determined by the temperature in a quite narrow density interval.

\section{Energetics of LB$_2$ states}

In this Section, we elucidate why the system appears to be in the
LB$_2$ phase. The energy functional (\ref{nuclenergy}) can be
rewritten in the form  \beq E[n] = 2\int\dfrac{p^2}{2m}n({\bf
p})\,d\upsilon + \dfrac{1}{2}\int f(\bq)S(\bq;[n])\,d\upsilon\,.\label{S(q)_enrnucl}\eeq Here the
interaction energy is given by means of the structure function
\beq S(\bq;[n]) = \dfrac{2}{\rho}\int n({\bf p}+{\bf q})n({\bf
p})\,d\upsilon\,.\label{structnucl}\eeq In the case $T=0$, which we address below,
the set of all possible QP momentum distributions
with one or two sheets of the Fermi surface (that is the FL or
LB$_2$ states) is specified by \bea &&n_2(\bp) =
\theta(p_2-p)-\theta(p_1-p)\,,\\&&\quad\qquad
p_2^3-p_1^3=p_F^3\,.\label{normT0}\eea The last equation follows
from the normalization condition (\ref{nuclnorm}). This class is
referred below as QP distributions of $n_2$ type. In this case, the
structure function is evaluated explicitly, the corresponding
formulas are given in the Appendix. It is sufficient, due to Eq.\ (\ref{normT0}),
to deal with one parameter. The convenient choice
is $\eta=p_2-p_1$ (the width of the occupied region in the QP
momentum distribution) that defines distribution parameters \beq
p_{1,2}=\dfrac{1}{\sqrt{3}}\sqrt{\dfrac{p_F^3}{\eta}-\left(\dfrac{\eta}{2}\right)^2}\mp\dfrac{\eta}{2}\,.\eeq
The value $\eta=p_F$ corresponds to $p_1=0,\,p_2=p_F$ i.e.\ to the case of the Fermi
step. The decrease of $\eta$ leads to a monotonic increase of
$p_{1,2}$.

\begin{figure}[]
\begin{minipage}[h]{0.49\linewidth}
\begin{center}
\includegraphics[width=1\linewidth,height=0.9\linewidth]{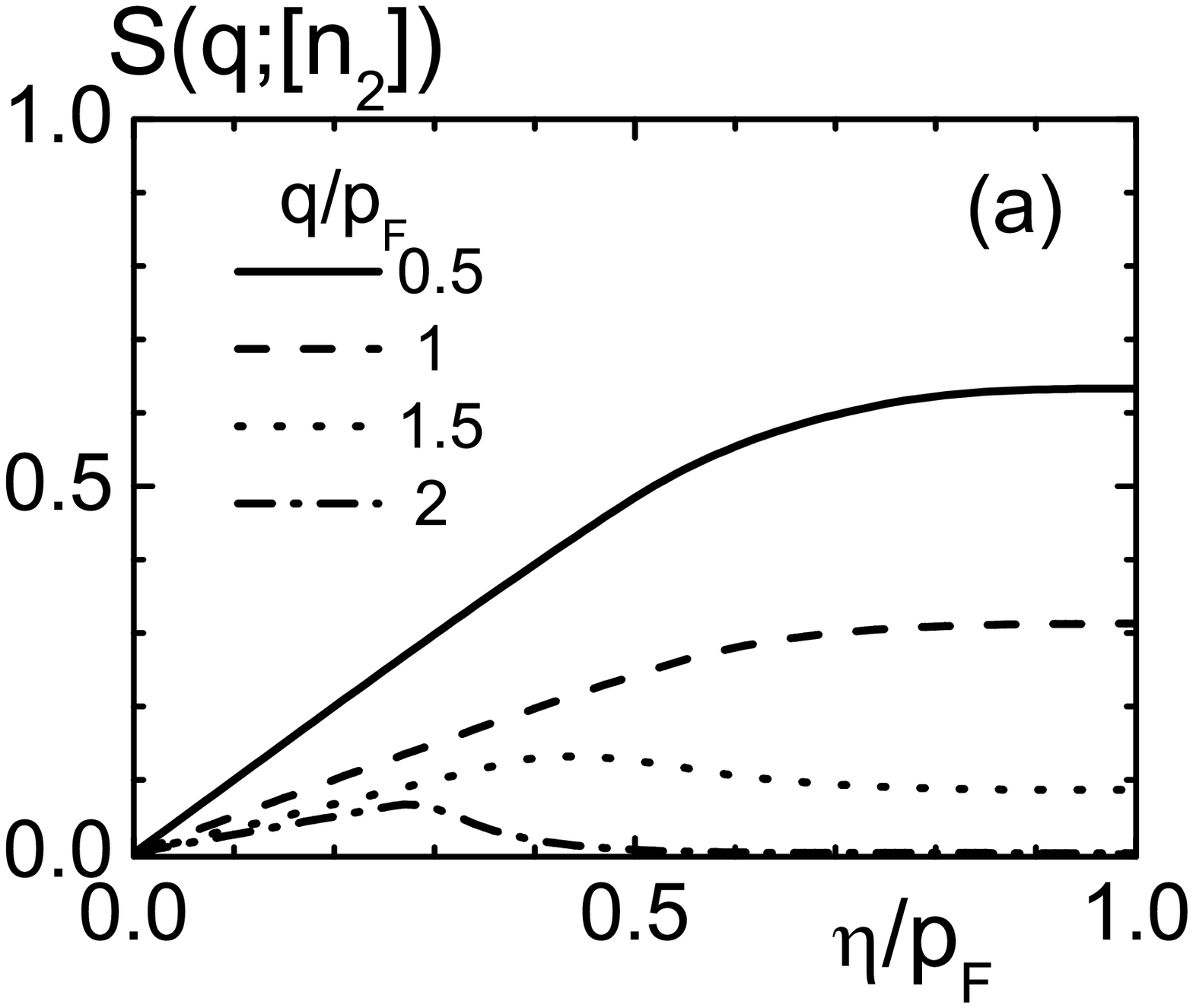}
\end{center}
\end{minipage}
\hfill
\begin{minipage}[h]{0.49\linewidth}
\begin{center}
\includegraphics[width=1\linewidth,height=0.9\linewidth]{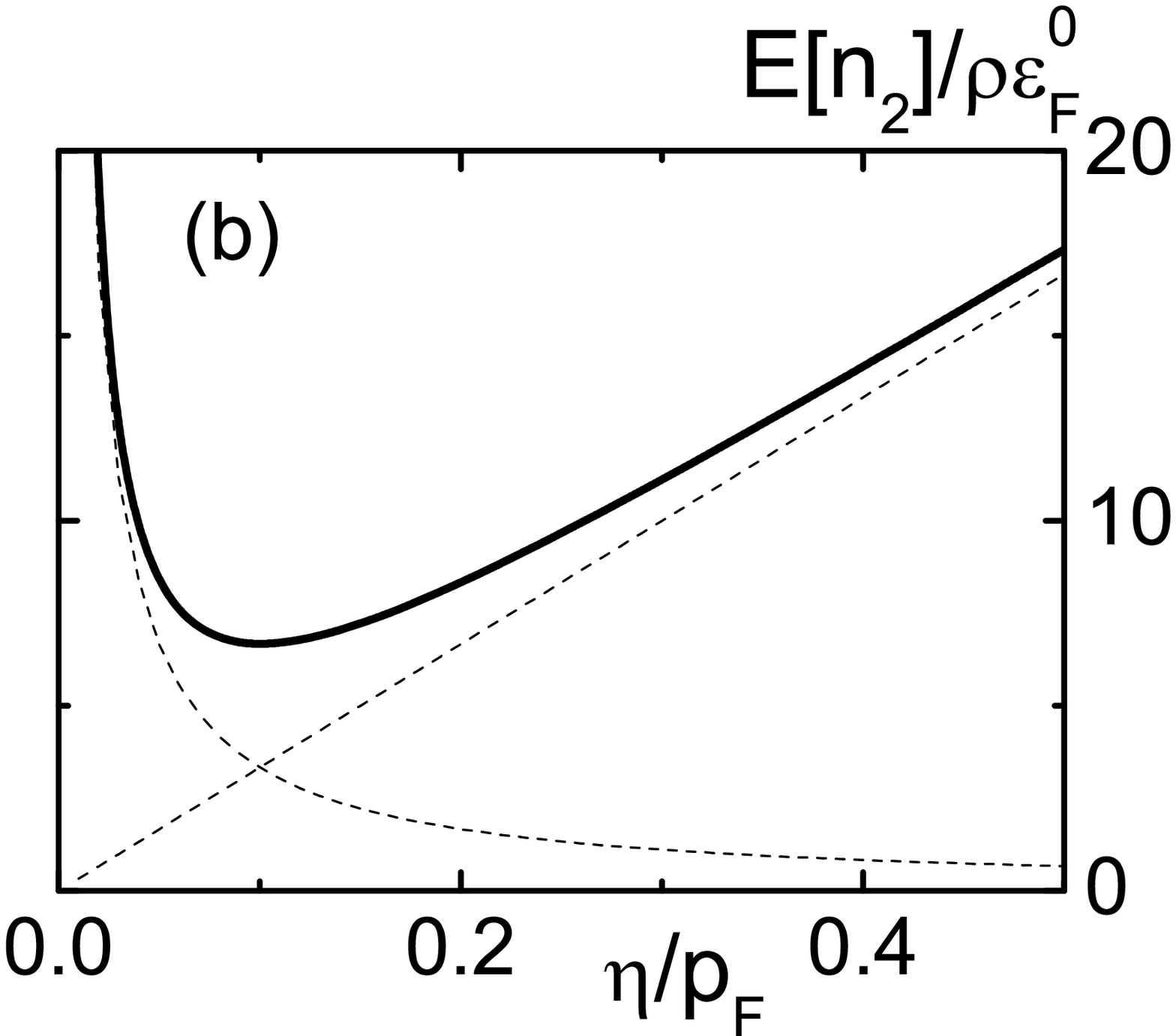}
\end{center}
\end{minipage}
\begin{center}
\begin{minipage}[c]{0.9\linewidth}
\caption{Left panel: The structure function for QP momentum distributions of the $n_2$ type versus the ratio $\eta/p_F$ at different values of the wave vector $q$. Right panel: The total energy per particle for the thin $n_2$ type momentum distributions.}\label{fig:SEH}
\end{minipage}
\end{center}
\end{figure}

The behavior of the structure function $S(\bq;[n_2])=S(q;\eta)$ is
shown in panel (a) of Fig.~\ref{fig:SEH}.
%
%

Its explicit form, corresponding to the Landau state ($\eta=p_F$), is
well known \begin{multline} S(q;p_F)=S_{\fl}(q)\\=\dfrac{1}{2}\left(1-\dfrac{q}{2p_F}\right)^2\left(2+\dfrac{q}{2p_F}\right)\theta(2p_F-q)\,.
\end{multline} It is remarkable that in the other important case of thin
($\eta\ll p_F$) $n_2$-type QP distributions that can appear by a
first-order TPT, one obtains
\beq\left.S(q;\eta)\right|_{\eta\rightarrow
0}=\dfrac{\eta}{2q}\,.\eeq This result is demonstrated well by
Fig.~\ref{fig:SEH}, and reflects a decrease of the overlap of the momentum distributions in the integral in Eq.~(\ref{structnucl}). As a consequence, this leads to a reduction of the interaction energy, \beq\left.E_{int}[n_2]\right|_{\eta\ll p_F}\simeq\dfrac{U\eta}{2p_F}\,,\quad U=p_F\int\dfrac{f(q)}{2q}d\upsilon\,.\eeq  The kinetic energy, on the contrary, shows enhancement in agreement with the explicit result \begin{multline} E_{kin}[n_2] =
\dfrac{3}{5p_F^5}\left(p_2^5-p_1^5\right)\eps^0_F\rho\\=\left(\dfrac{p_F}{3\eta}+\dfrac{\eta^2}{3p_F^2}-\dfrac{\eta^5}{15p_F^5}\right)\eps^0_F\rho\,.\end{multline}
The total energy in the case of distributions of the $n_2$ type in the limit
$\eta\ll p_F$ is given by \beq E[n_2] \simeq
\left(\dfrac{p_F}{3\eta}+\dfrac{u\eta}{2p_F}\right)\eps^0_F\rho\,,\quad
u=U/\eps^0_F\rho\,.\label{totEmod}\eeq This function is plotted on panel (b) of
Fig.~\ref{fig:SEH} and shows a minimum value \beq E_{\lb} =
\sqrt{\dfrac{2}{3}u}\;\eps^0_F\rho\label{En2min}\eeq at the point
$\eta_c=p_F\sqrt{2/(3u)}\,.$ Finally, we note that the
$E_{\lb}$ energy appears to be lower than the Fermi-liquid one,
\beq E_{\fl} = \dfrac{3}{5}\eps^0_F\rho + \dfrac{1}{2}\int
f(q)S_{\fl}(q)\,d\upsilon\,,\label{EnFL}\eeq when the QP
interaction is sufficiently strong. Indeed, if one characterizes
the QP interaction function by an effective coupling constant
$f(q)\propto g$, Eqs.\ (\ref{En2min}) and (\ref{EnFL}) show that $E_{\lb}\propto\sqrt{g}$ and $ E_{\fl}\propto g$ at the large-$g$ limit.

Thus, the appearance of LB$_2$ states is explained by the
interplay between the kinetic and the interaction energy
contributions. The existence of this energy minimum, generally,
does not dependent on whether the Landau state is present or not.
We remark that this simple explanation is limited to regarding the $n_2$ set
of QP momentum distributions. A generalization to a more complete
class $n_{\alpha>2}$ could in principle reveal an instability of a
LB$_2$ state with respect to some energetically favored one
\cite{PanZver2012}. The solid proof of the existence of the LB$_2$
ground state comes from the direct solution of Eqs.\ (\ref{eqn(p,T)}),(\ref{nuclnorm}), and (\ref{eqeps}).

\section{Conclusion}

In this article we have considered two scenarios of topological
phase transitions in homogeneous neutron matter. The transitions
occur between the Fermi-liquid state and the other one with two
sheets of the Fermi surface. The investigation was performed with
the use of a semi-microscopic expression for the quasiparticle
interaction function in the vicinity of $\pi^0$ condensation
point. The order of the phase transition is shown to depend on
the value of the critical wave vector $q_c$. The first
possibility for a rearrangement of the quasiparticle degrees of
freedom is the second-order topological phase transition. It
occurs when $q_c<p_F$ and corresponds to a quantum-critical-point scenario \cite{KhodelTwoScen2007,KhodelTopCros2011,KhodelQCP2008} of the
Fermi surface reconstruction. The second possibility studied in
this work is the first-order topological phase transition. This case occurs when
$q_c\gtrsim p_F$ and is connected with a
sudden change in the quasiparticle momentum distribution and
spectrum. The first-order topological phase transition can be
driven by change of either the density or the temperature.
Thermodynamic functions and the phase diagram have been calculated. It
is shown that the influence of the temperature on the Fermi
surface topology is essential in a quite narrow density region.
A simple explanation of the origin of the first-order topological
phase transition at $T=0$ is given.

\section{Acknowledgments}

We thank V.~A.~Khodel and E.~E.~Saperstein for their interest to
this work and useful discussions. One of the authors (S.S.P.)
would like to thank INFN (Sezione di Catania) for hospitality
during his stay in Catania. This research was partially supported
by Grants No.\ NSh-7235.2010.2 and No.\ 2.1.1/4540 of the Russian
Ministry for Science and Education, and by the RFBR Grants
No.\ 11-02-00467-a and No.\ 12-02-00955-a.

\section{Appendix}

The structure function within the $n_2$ set of QP momentum
distributions reads \beq S(\bq;[n_2]) = \dfrac{2}{\rho}\int n({\bf
p}+{\bf q})n({\bf p})\,d\upsilon = S_{11} + S_{22} -
2S_{12}\,.\eeq Here $(4\pi/3)p_F^3\,S_{ij}$ is a volume the intersection of two
spheres with radii $p_{\min} = p_i$ and $p_{\max} = p_j$, $p_{\min}\le p_{\max}$, while the distance between their centers is equal to the $\bq$ vector length. The value of the volume is
specified by the expression \begin{multline} S_{ij}(q;p_i,p_j) = \bigl(\zeta(q;p_i,p_j)+\zeta(q;p_j,p_i)\bigr)\\\times\theta(p_i+p_j-q)\theta(q+p_i-p_j)
+p_i^3\theta(p_j-q-p_i)\,.\label{Spp}\end{multline} The first term
corresponds to an intersection case, while the second one corresponds to a
complete enclosure of the smaller sphere into the bigger one. The
function $\zeta$ has the form \begin{multline} \zeta(q;p_i,p_j) =\dfrac{1}{4}\left(p_i-\dfrac{p_i^2+q^2-p_j^2}{2q}\right)^2\\
\times\left(2p_i+\dfrac{p_i^2+q^2-p_j^2}{2q}\right).\label{Sffun}\end{multline}

\end{document}